\documentclass[twoside]{article}
\usepackage{qic}
\usepackage{graphicx}
\usepackage{psfig}
\usepackage{exscale}
\usepackage{amsmath}
\usepackage{latexsym}
\usepackage{amsfonts}
\newcommand{\Tr}{\mbox{ Tr }}
\newcommand{\floor}{\mbox{ floor }}
\newcommand{\eins}{\mbox{$1 \hspace{-1.0mm}  {\bf l}$}}
\newcommand{\ket}[1]{ | \, #1  \rangle}
\newcommand{\bra}[1]{ \langle #1 \,  |}
\newcommand{\calp}{\mbox{$\mathcal P$}}
\newcommand{\cala}{{\mathcal S}_1}
\newcommand{\calb}{{\mathcal S}_2}
\newcommand{\calpw}{{\mathcal P}_W^{\cala}} 
\newcommand{\calab}{\mbox{$\mathcal H_A \otimes \mathcal H_B$}}
\newcommand{\botimes}{ \mbox{\large $\otimes$} }
\newcommand{\field}{ {\mathcal K} }
\newcommand{\calAB}[1]{\mbox{$({\mathcal H}_A \otimes \field^{#1})
  \botimes ({\mathcal H}_B \otimes \field^{ #1})$}}

\begin{document} 

\fpage{1}

\centerline{\bf
SIMPLIFYING  SCHMIDT NUMBER WITNESSES VIA}
\vspace*{0.035truein}
\centerline{\bf HIGHER-DIMENSIONAL EMBEDDINGS}
\vspace*{0.37truein}

\centerline{\footnotesize
FLORIAN HULPKE, DAGMAR BRUSS, MACIEJ LEWENSTEIN and ANNA SANPERA}
\vspace*{0.015truein}
\centerline{\footnotesize\it 
Institut f\"ur Theoretische Physik,
Universit\"at Hannover} 
\baselineskip=10pt
\centerline{\footnotesize\it D-30167 Hannover, Germany}
\vspace*{0.225truein}

%\publisher{(received date)}{(revised date)}

\abstracts{
We apply the generalised concept of witness operators to 
arbitrary convex sets, and review 
the  criteria for the optimisation of these
general witnesses.
We then define an embedding of  state vectors 
and operators into a higher-dimensional
Hilbert space. This embedding leads to a 
connection between any Schmidt number
witness in the original Hilbert space and a  witness
for Schmidt number two (i.e. the most general entanglement witness) in 
the appropriate enlarged Hilbert space. 
Using this relation we arrive at a 
conceptually simple method for the construction of 
Schmidt number witnesses in bipartite systems.
}{}{}

\vspace*{10pt}

\vspace*{3pt}
%\communicate{to be filled by the Editorial}

%\pacs{03.67.-a, 03.65.Ud, 03.67.Mn}

\vspace*{1pt}\textlineskip    
   
\section{Introduction} 
In spite of the tremendous effort devoted in the recent years 
to characterize  (i.e. to classify, quantify, detect and measure)
entanglement
\cite{entanglement}, the description of entanglement 
remains an eluding problem whose
complexity grows very fast with the number of subsystems of a given
composite quantum system 
and with the dimension of the Hilbert space.
Several operational 
separability criteria have been introduced 
to determine if a  given state $\rho$ acting on $ 
{\cal H}={\cal H}_1 \otimes {\cal H}_2\otimes ... \otimes {\cal H}_n $
is entangled or not (i.e separable).
Among them,  the criterion of the 
Positive Partial Transposition (PPT)  \cite{peres-horodecki}
and the realignment criterion \cite{realignment}
are particularly powerful. Recently Doherty et al. \cite{doherty}
have introduced a new family of separability criteria 
which gives a characterization of mixed bipartite entangled states with a finite number of tests.
 
An apparently different approach to treat the same problem 
is based on entanglement witness operators \cite{horodecki-terhal,lkch}. 
An entanglement  witness  $W$ is a hermitian 
operator which has a positive expectation value  on all 
separable states, but a negative one for at least one entangled state. 
This state is said to be detected by the 
witness operator. The existence of these operators is a direct
consequence of the nested convex structure of the sets of (mixed) states 
acting on the Hilbert space  ${\cal H}$ of a composite system. 
Since for any given (finite-dimensional) Hilbert space
the subset of separable states is  convex and closed, 
it is always possible to find entanglement witness operators 
regardless of the dimensions and/or the number of subsystems 
of the composite system. 
Equivalently each entangled state can be detected 
by a witness operator. Notice that
by using this approach  the problem of determining 
whether a given state $\rho$ is entangled or not 
is transformed into the problem of
{\it finding} a suitable witness operator that detects it. The most
suitable witnesses will be those that detect more states than any
other ones, and for that reason they are called optimal witnesses. 

Remarkably, entanglement witnesses have become a very powerful 
tool not only for detecting entanglement, 
but also in the context of various 
other tasks in quantum information theory.
For instance, establishing a secret key in quantum cryptography requires the 
existence of quantum correlations, which can be characterized by optimal 
witnesses \cite{curty}. The activation and
distillation properties of a state $\rho$ (that is, the possibility
to locally distill from an ensemble of mixed states a subset of maximally  
entangled pure states) can also be recast in terms of witness operators
\cite{kraus}. By far the best-known and most famous 
entanglement witnesses are the so-called Bell inequalities \cite{Terhal}.
It is easy to see from the definition of entanglement witnesses
that each Bell inequality corresponds to an entanglement witness. However,
this correspondence does not necessarily hold the other way round, as
there exist entangled states that do not violate any Bell inequality but
nevertheless are detected by a witness operator \cite{Werner}. 
Thus, for the detection of entanglement witness operators 
are, in this respect, {\em stronger} than Bell inequalities. 
Furthermore, witness operators can be 
generalized to distinguish between 
different {\em types} of entanglement as long as the different entanglement 
types correspond to   
nested convex subsets. This is indeed the case for bipartite systems 
and, at least, for the simplest 
multipartite system, i.e. ${\cal H}={\cal C}^2 \otimes {\cal C}^2
\otimes {\cal C}^2$ \cite{acin}. 
For larger multipartite systems, where
the structure of entangled states is much richer
and  much less-known, witness operators are also a 
useful tool to  explore the  structure of the Hilbert space.
Finally, let us point out that since entanglement witnesses 
are observables (although not positive semidefinite) they can be
measured. The experimental
implementation of a witness operator can be realised by means of {\em local}
measurements \cite{ottfried,pittenger}
and has already been achieved in the laboratory \cite{demartini,weinfurter}.

The paper is organized as follows.
In section \ref{Sect2} we first review the concept of a general witness as well
as its optimization following the arguments given in 
\cite{lkch,lkhc, reflections}.
Most of the lemmas and theorems stated in this part of the paper are 
a straightforward generalization of the 
formalism developed previously, but for  completeness we have
 included them here. 
In section \ref{Sect3} we restrict ourselves to bipartite systems, 
and review first the concepts of Schmidt number and Schmidt number witnesses.  
We show then that by embedding the state vectors 
and operators of the original Hilbert space into a higher-dimensional
Hilbert space it is possible to connect any Schmidt number
witness in the original Hilbert space to a witness of Schmidt number two (i.e. the most general entanglement witness) in 
the appropriate enlarged Hilbert space. 
Using this method one can 
simplify the construction and optimization of the desired general witness. 
We close this section with an explicit 
example to illustrate our method.  
Finally,  we present our conclusions in section \ref{Sect4}.
 
\vspace*{1pt}\textlineskip    
\section{Optimisation of a general witness operator}
\label{Sect2}

We consider quantum systems of arbitrary, finite dimensions. 
By ${\mathcal H}$ we denote a Hilbert space over the field $\field$,
by 
${\mathcal B}({\mathcal H})$ the space of bounded operators 
acting on this Hilbert space 
and by ${\mathcal P} \subset 
{\mathcal B}({\mathcal H})$ the subset of positive semidefinite operators 
with trace one (the set of states on ${\mathcal H}$).
\begin{figure}
\begin{minipage}[c]{.45\textwidth}
\centerline{\includegraphics[width=.6\textwidth]{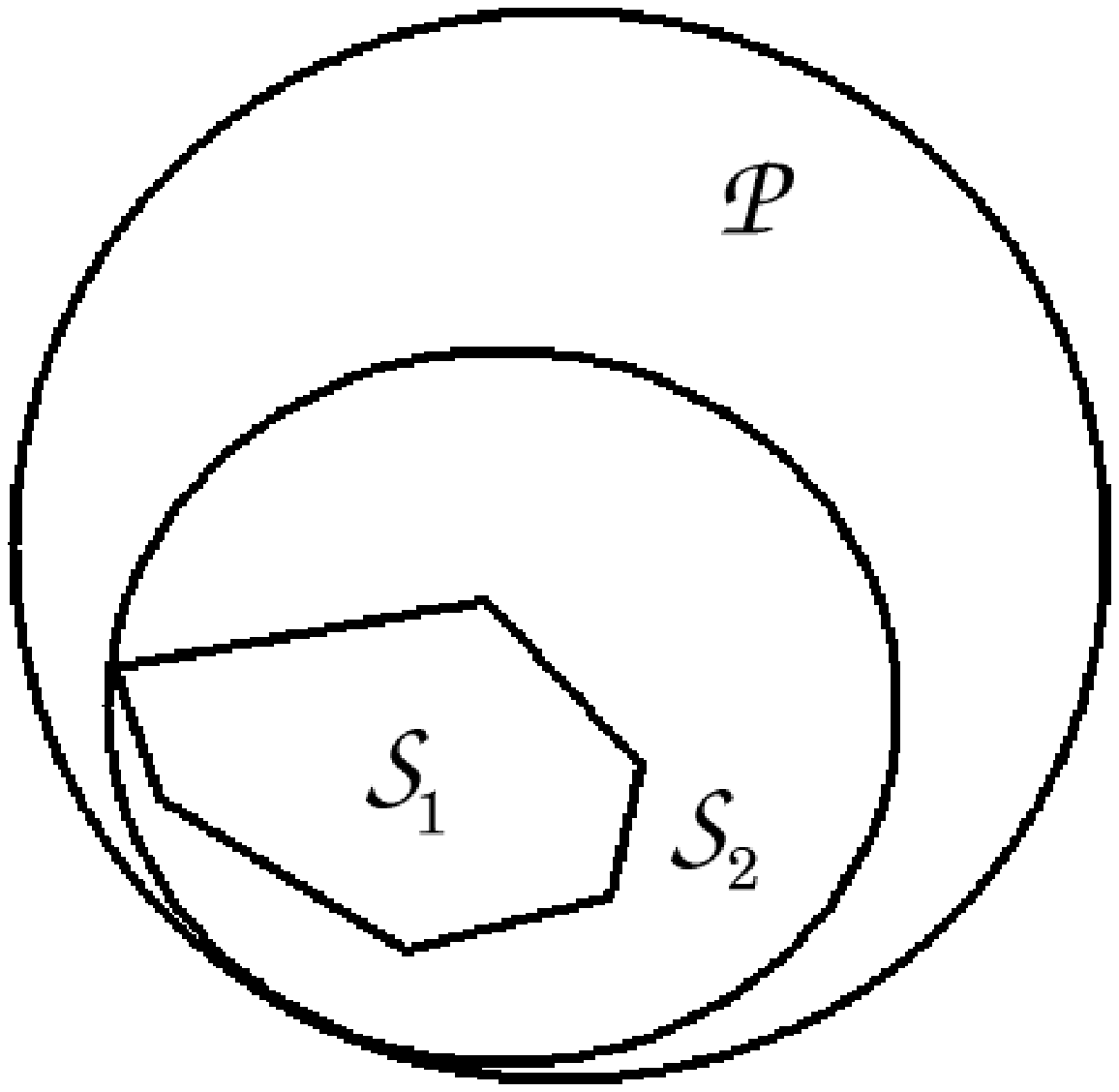}}
\vspace*{13pt}
\fcaption{\label{fig1}The structure of the nested convex sets.}
\end{minipage}
\begin{minipage}[c]{.45\textwidth}
\includegraphics[width=.8\textwidth]{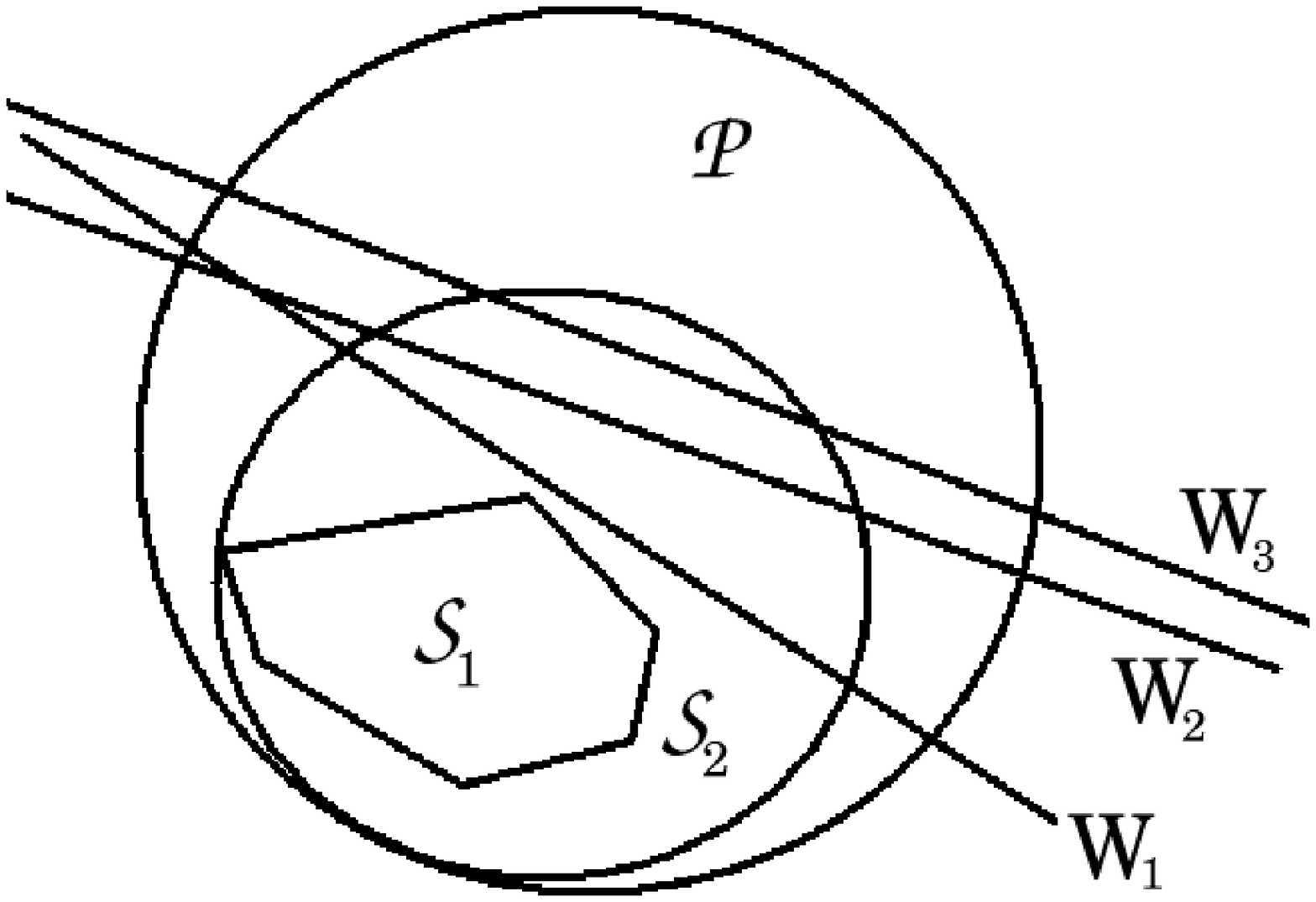}
\vspace*{13pt}
\fcaption{\label{fig2}The witness corresponding to the hyper-plane  $W_1$\\
   is
  (${\mathcal S}_1,{\mathcal S}_2$)-finer, but not (${\mathcal S}_1,{\mathcal
  P}$)-finer than $W_2$. 
  Both $W_1$ \\
and $W_2$
are (${\mathcal S}_1,{\mathcal P}$)-finer than $W_3$.}
\end{minipage}
\end{figure}

Consider the following situation, sketched in figure \ref{fig1}: for
two given nested convex, 
closed subsets $\cala \subset \calb \subset {\mathcal P}$, does 
a given $\rho \in \calb$ belong to $\cala$?
Without loss of generality we will assume that the identity belongs
to $\cala$ and that $\cala$ is not of measure zero in $\calb$
\cite{measurezero}. 

\vspace*{12pt}
\noindent {\bf Definition 1:}
For any convex set $X$ we denote the {\it border of X} by
\begin{equation}
\delta X := \left\{ a \in X \left|
\begin{array}{c}  
\exists b \in X,\; s.t.\;\; \forall \lambda > 0\;\; \mbox{ one has}:\\
(1+\lambda) a - \lambda b \not\in X
\end{array}
\right.
\right\}
\end{equation}
Some of the operators on the border, that have special properties, 
are  referred to as edge-operators. 
Notice that a full characterization of the border-operators  of $\cala$ 
immediately implies a full characterization of the operators that belong to
$\calb\backslash\cala$. In fact a full characterization of 
edge-operators is sufficient for this aim.  
Without loss of generality one can always 
shift the set $\cala$ such that  
$\eins \in \cala \backslash \delta \cala$.

Before answering the question  whether
a given $\rho \in \calb$ belongs to $\cala$
we first  formally define  a general witness
operator. To simplify the notation,
by $W$ we shall denote
an (${\cala ,\calb}$)-witness  defined as follows:

\vspace*{12pt}
\noindent {\bf Definition  2:} 
A hermitian operator $W$ is called an $(\cala, \calb)$-witness iff:\\
(i) $\Tr(W \sigma) \geq 0\;\;\; \forall \sigma\in \cala$.\\
(ii) $\exists \rho\in \calb$ such that $\Tr( W \rho ) <0$. \\
(iii) $\Tr W =1$.\\
We note in passing that $W$ is not positive semidefinite  and that 
condition (iii) corresponds to a normalization of the operator
$W$. This normalization is useful for the comparison between  
different witnesses. 
\\ 
To prove that $\rho \in \calb \backslash \cala$ it 
is sufficient to find a witness operator $W$ that detects $\rho$. 
For the cases in which the set $\cala$ is also closed  
(and therefore compact, due to the boundedness of $\cal P$) 
the existence of a witness that detects $\rho$ is also necessary
for $\rho \in \calb \backslash \cala$
\cite{necess}.

Note that without the requirement
$\eins \in \cala$ there could be cases for
which no witnesses exist. For example,  assume $\sigma
\not = \eins$, $\cala=\{\sigma\}$ and 
$\calb=\{ \rho(\lambda)=\lambda \eins+ (1-\lambda) \sigma | 1 \geq \lambda 
>
0 \}$.  Then no linear operator exists such that $\Tr (W \sigma) \geq
0$ for all $\sigma\in \cala$ and  $\Tr (W \rho) <0$ for a $\rho \in
\calb$ and $\Tr W=1$.\\ 

To proceed further, let us introduce now some basic concepts and
notations related to witness operators. We shall adopt here the
notation developed in Ref. \cite{lkch} and \cite{reflections}.
Our notation is as follows: \\ 
1. $D_{W}^{\calb}:=\{ \rho\in {\calb} | Tr (W \rho) <0 \}$,
i.e. the set of the states in ${\calb}$ that are detected by $W$.\\ 
2. ${\cal Q}_{\calb}:= \{ Z | \Tr (Z \rho) \geq 0 \mbox{ } \forall \rho \in
{\calb} \}$, i.e. the set of operators which do not detect any state in $
{\calb} $, and, therefore, are non-negative on $\calb$.\\ 
3. $\calpw:= \{ | \psi \rangle \langle \psi| \in {\cala} | \langle
\psi|W|\psi\rangle =0 \}$, i.e.
 the set of one-dimensional projectors (pure states) 
in ${\cala}$ for which the expectation value of $W$ vanishes.\\ 
4. Finer witness: $W_1$ is $(\cala, \calb)$-finer than $W_2$ iff
$D_{W_2}^{\calb} \subset D_{W_1}^{\calb}$, that is, if any state
detected by $W_2$ is also detected by $W_1$. \\ 
5. Optimal witness:
$W$ is an $(\cala, \calb)$-optimal witness iff there exists no other 
witness that is $(\cala,\calb)$-finer than
$W$.  So for all $W_1$ that are $(\cala, \calb)$-finer than $W$ the 
equality $W_1=W$
holds.\\
\\
The concept of a witness being $(\cala, \calb)$-finer than another one
depends on $\calb$.
 As illustrated in figure
\ref{fig2} it is possible that a witness $W_1$ is
($\cala,\calb$)-finer than $W_2$, but that there exists
 an $\calb'$ with
 $\calb \subset \calb'$, such that  $W_1$
is not $(\cala,\calb')$-finer.

\noindent The above definitions provide the necessary 
tools to optimize a given general witness.

\vspace*{12pt}
\noindent{\bf Lemma 1 (Lemma 1 in \cite{lkch}):} Let $W_1$ be $(\cala,\calb)$-finer than
$W_2$ and
\begin{equation}
\lambda := \inf_{\rho \in D_{W_2}^{\calb}} 
\frac{ \Tr \rho W_1 }{\Tr \rho W_2}
\end{equation}
Then for any positive operator $\rho$ we have:\\
(i) If $\Tr \rho W_2=0$ 
then $\Tr \rho W_1\leq 0$.\\
(ii) If $\Tr \rho W_2<0$ then $\Tr \rho W_1 \leq
\Tr \rho W_2$.\\
(iii) If $\Tr \rho W_2>0$ then $\Tr \rho W_1 \leq
\lambda \Tr \rho W_2$.\\
(iv) $\lambda \geq 1$. In particular if $\lambda=1 \Leftrightarrow W_1 =W_2$.\\
\\
In complete analogy with the optimization of entanglement witnesses, 
one can also construct $(\cala,\calb)$-finer general witnesses 
by simply subtracting any operator that is positive definite 
on $\calb$ (i.e. $Z \in {\cal Q}_{\calb}$) from 
the original witness in such a way that the remaining operator 
still fulfills the necessary conditions for being a witness operator.

\vspace*{12pt}
\noindent{\bf Lemma 2 (Lemma 2 in \cite{lkch}):}
$W_1$ is $(\cala,\calb)$-finer than $W_2 \Leftrightarrow$  there exists a 
$Z \in {\cal Q}_{\calb}$ (i.e. $\Tr(Z \rho) \geq 0\;$ for all $\rho \in
{\calb})$, and 
there exists an $\epsilon$ with $1 >\epsilon\geq 0$ such
that $W_2= (1-\epsilon) W_1 + \epsilon Z$ 
$\left(\Leftrightarrow \right.$
 for all $\left. \rho \in \cala: \Tr(W_2 \rho) -
(1-\epsilon) \Tr(W_1 \rho) \geq 0 \right)$.

\vspace*{12pt}
\noindent {\bf Theorem 1 (Theorem 1 in \cite{lkch}):}
A witness $W_1$ is optimal $\Leftrightarrow$ 
for all $Z\in {\cal Q}_{\calb}$ and $\epsilon>0$ the operator
$W':=(1+\epsilon) W_1 - \epsilon Z$ is not a witness, i.e. it  
does not fulfill $\Tr(W' \rho)\geq 0$ for all $\rho \in {\cala}$.

\vspace*{12pt}
\noindent {\bf Lemma 3( Lemma 3 in\cite{lkch}):} 
If $Z\calpw \neq 0$ (i.e.
 for all $P\in \calpw$ one has $Z \not\perp P$),
 then $Z$ cannot be subtracted from W, that is, 
$(1+\epsilon)W - \epsilon Z$ is not a witness for any $\epsilon >0$.

\vspace*{12pt}
\noindent {\bf Corollary 1 (Corollary 2  in \cite{lkch}):} 
If $\calpw$ spans ${\cal H}$ then W is optimal.

\vspace*{12pt}
\noindent {\bf Lemma 4:} 
If, for a given $(\cala,\calb)$-witness $W$, there exists a
$\rho \in \delta \cala \backslash \delta {\calb}$,
such that $\Tr W\rho =0$, then $W$ is $(\cala,\calb)$-optimal.
Geometrically  $W$ can be interpreted as 
an $\cala$-edge witness operator, namely it is tangent to the set $\cala$
at a point which does not belong to the border of $\calb$.

\vspace*{12pt}
\noindent{\bf Proof:}
Assuming that $W$ is not $(\cala,\calb)$-optimal, there would exist an
$(\cala,\calb)$-finer general witness $W'\not= W$.
Following the assumption there
has to exist a state $\sigma \in \calb$ that is
detected by $W'$ (i.e. $\Tr W' \sigma <0$) but not by $W$ (i.e. $\Tr W
\sigma \geq 0$). Since $\rho \in \cala$ and $W'$ is $(\cala,\calb)$-finer
than $W$, we know according to Lemma 1 (i) that $\Tr W' \rho=0$.
Furthermore  $\rho \not \in \delta \calb$.
This leads by
definition to the fact that there is a $\lambda > 0$, with the
property that
the state $(1+\lambda)\rho-\lambda \sigma
\in \calb$. Evaluating
both witnesses $W$ and $W'$ on this state
we see that $\Tr W ( (1+\lambda)\rho-\lambda \sigma )
= - \lambda \Tr W \sigma \leq 0$, but
$\Tr (W'(1+\lambda)\rho-\lambda \sigma ) = -
\lambda \Tr W' \sigma > 0$. 
This is in contradiction to Lemma 1 (ii). 
Therefore, we conclude that there exists no witness $W'\not= W$ 
which is $(\cala,\calb)$-finer than $W$, 
and since all witnesses which are
$(\cala,\calb)$-finer than $W$ have to be equal to $W$, we conclude that
$W$ is optimal $\square$\,.

Witness operators are often used
to detect generic entanglement in bipartite systems.
In this case the two convex sets 
$\cala, \calb$
can be identified as 
$\cala={\mathcal S}=\{\mbox{set of all separable states} \}$ and 
$\calb={\mathcal P}=\{ \mbox{set of all positive (semidefinite)
operators with trace one} \}$. 
It is well known that for
systems acting on a Hilbert space of dimension $\dim{\mathcal H}>6$
(where $\dim{\mathcal H}_A, \dim{\mathcal H}_B\geq 2$)
there exists 
entanglement which cannot be detected by means of the partial
transposition. The set of states that remain positive
under partial transposition (PPT-set) also form a convex set. 
We can distinguish  PPT-entangled states
from separable ones if we search for witnesses  associated to 
$\cala={\mathcal S}$ and 
$\calb=\{ {\mathcal{PPT}} \}$
with $\cala \subset \calb$. 
Those witnesses $W$ which are
 capable to detect $\rho\in \calb\backslash\cala$ are
necessarily non-decomposable, since they cannot be written as $W=P+Q^T$,
 where $P$ and $Q$
are positive semidefinite operators \cite{lkch}. 
First examples of non-decomposable entanglement witnesses were
provided in~\cite{Terhal}, and the characterisation of such witnesses was
presented in~\cite{lkhc}. Notice that the
concept of witness operators only relies on a  nested subset structure, 
and thus is not restricted to bipartite systems.  
For multipartite systems, there exist typically various classes of 
distinct multipartite entanglement.
For instance, for $2\times 2\times 2$ systems, there are classes of
separable $\mathcal S$, 
biseparable $\mathcal B$,  $\mathcal W$ \cite{duer} and 
$\mathcal {GHZ}$ mixed states, which are ordered in the nested structure
$\mathcal S \subset  \mathcal B  \subset \mathcal W\subset \mathcal {GHZ}$
\cite{acin}. For this case 
and similar ones one can tailor the appropriate  
witness $W$ to discriminate between the different nested convex sets.  

In the next section we will study  so-called Schmidt (number)
 witnesses \cite{schmidtwit}, which
 detect how many degrees of freedom
of the subsystems of a bipartite system are entangled
with each other. Thus Schmidt witnesses
can distinguish between entangled states from
 different Schmidt classes. In this
sense Schmidt number witnesses provide a refinement of general entanglement
witnesses.

\vspace*{1pt}\textlineskip    
\section{From Schmidt number witnesses to entanglement witnesses}
\label{Sect3}

For bipartite systems it is possible to extend the useful 
concept of the Schmidt {\em  rank} for pure states \cite{Schmidt} 
to the Schmidt {\em number} for mixed
states  \cite{TerhalHoroSchmidt}. 
The Schmidt number of a mixed state $\rho$ 
characterises the maximal
Schmidt rank of the pure states which is {\em at least} needed to construct
$\rho$. It 
is defined as:
\begin{equation}
SN(\rho ):= \min_{\rho=\sum_{i=1}^r p_i |\psi_i \rangle \langle
  \psi_i|} 
\max_{i}\{ SR(\psi_i)\},
\end{equation}
where  $SR(\psi_i)$ is the Schmidt rank of  $\ket{\psi_i}$.
By $(SN)_k$ we denote the 
Schmidt class $k$, i.e. the
set of states that have  Schmidt number 
$k$ or less.
This set is a convex, compact subset 
of $\calp$, and there is a nested structure of the form 
$(SN)_1\subset (SN)_2\subset...\subset(SN)_k\subset...\subset {\cal P}$. 
Clearly, $(SN)_1=\cal{S}$ corresponds to the set of separable states. 
As explained in \ref{Sect2} it is clear that one can construct a Schmidt number
witness operator $S_k$ which is non-negative on
all states in $(SN)_{k-1}$, but detects at least one state belonging to $(SN)_k$.
Using our previous notation this would correspond to a 
$\left( (SN)_{k-1},(SN)_k\right)$-witness, but as a shorter
notation and to be consistent with the notation of \cite{schmidtwit} we
denote $S_k$ as a {\em $k$-Schmidt-witness} ($k$-SW). In the same way we define
the terms {\em $k$-finer} and {\em $k$-optimal} as abbreviations.

In  the previous part of this paper we have established tools to answer
 (a)
under
which conditions  a given $k$-SW is   $k$-finer than another one, and
(b) how to optimise a given $k$-SW (namely one has to subtract operators
$Z$ which are positive definite on the set $(SN)_{k}$,  
 such that the operator $W=S_k-Z$ has the property
$\Tr(W \sigma)\geq 0$  
for all $\sigma \in (SN)_{k-1}$).
To verify 
positivity on all states from the set $(SN)_{k-1}$
 it suffices to restrict oneself
to the set of all pure states $\ket{\psi_{k-1}}$,
 since those
are the extremal points of  $(SN)_{k-1}$. 
But nonetheless such  verification remains laborious, 
since these states  
do not exhibit a particularly useful structure that allows  to
check easily whether $\Tr(W \ket {\psi_{k-1}}\bra{\psi_{k-1}})\geq 0\;$ 
for all $\ket {\psi_{k-1}}$.
Notice, however, that for the case of a general entanglement witnesses (or more 
precisely, of 2-SW's)
this problem becomes much easier, since 
semi-positivity of an operator $W$ on all  product states
 $|e,f\rangle \in 
\calab$ 
is equivalent to the semi-positivity of the $(\dim {\mathcal
  H}_A-1)$-parameter 
family of
operators  $\langle e|W|e\rangle 
\in {\mathcal B}({\mathcal H}_B)$. 
Thus, rather than checking a
\mbox{$(\dim {\mathcal H}_A-1)$}$(\dim {\mathcal H}_B-1)$-dimensional 
parameter space corresponding to all product vectors, the task
is greatly simplified. Furthermore, 
positivity on a whole space  is 
mathematically a simpler concept than positivity on a given subset. 

Let us point out for  clarity that {\em every} Schmidt number witness is an entanglement
witness since it detects some kind of entanglement.  A 
witness for Schmidt number two  (2-SW) corresponds to a
witness that detects {\em all}
 kinds of entanglement without discriminating between the different types. Therefore
we sometimes use the name ``entanglement
witness''  synonymously with ``2-SW''.

In this section we will show that it is always possible to reformulate 
the problem of finding a witness of Schmidt class $k$ 
into  finding a $2$-SW denoted by ${\mathbb S}$ in a higher-dimensional
Hilbert space. Our method relies on  
embedding the original Hilbert space  ${\cal{H}}_A\otimes {\cal{H}}_B $ 
into an enlarged Hilbert space, such that
each pure state with Schmidt rank $k$ 
or less in the original space becomes a 
product state in the enlarged space. 
The enlarged Hilbert space
consists of the original one with two added local ancillas 
of dimension $k$, i.e.
${\cal H}_{\mbox{enlarged}}= \calAB{k}$.
The embedding is performed by means of a map $I_k$. We shall study
 the effect of this map on the set of operators acting on  
$\calab$ and show that this map also connects the expectation values 
of operators in the original Hilbert space with the expectation values 
of  operators in the enlarged Hilbert space. 

\subsection{Mapping from the original Hilbert space to the enlarged 
one}
In this subsection we first
define  a map which transforms states in the
original Hilbert space $\calab$ into states in the enlarged one. We 
then study 
the effect of this map on the operators acting on $\calab$ and define for
each operator $S$ an operator $ \mathbb S$ acting in the enlarged space.

\vspace*{12pt}
\noindent{\bf Definition 3:}
We denote by $ I_{k}$:  ${\cal H}_{A}\otimes {\cal H}_{B}\rightarrow 
\calAB{k}$ a map that transforms every pure
state $|\psi \rangle\in \calab $ into 
a pure state $\ket {I_k(\psi)}\in \calAB{k}$.
This map is defined by:
\begin{eqnarray}
|\psi \rangle= \sum_{i=1}^n \lambda_i 
|a_i b_i \rangle \mapsto |I_{k}(\psi) \rangle &=&
\left( \sum_{i=1}^{k} |a_i \rangle \otimes | {\mathbf i} \rangle \right) 
\otimes 
\left( \sum_{j=1}^{k}  \lambda_j |b_j \rangle \otimes |\mathbf{j} \rangle 
\right) \nonumber \\
&+&
\left( \sum_{i=k+1}^{2k} 
|a_i \rangle \otimes |\mathbf{i-k} \rangle  \right) \otimes 
\left( \sum_{j=k+1}^{2k}  \lambda_j |b_j \rangle \otimes |\mathbf{j-k} \rangle 
\right) + ... \nonumber \\
&+&
\left(\sum_{i=u+1}^{n} 
|a_i \rangle \otimes |\mathbf{i-u} \rangle \right) \otimes 
\left( \sum_{j=u+1}^{n} 
\lambda_j |b_j \rangle \otimes |\mathbf{ j- u} \rangle 
\right).
\end{eqnarray}
where $n\le$ min(${\mbox{dim}}{\cal H}_{A}, {\mbox{dim}}{\cal H}_{B})$.  
Here we have fixed a  basis
$\{|{\mathbf 1} \rangle
, ... , | {\mathbf k} \rangle \}$ for both ancilla spaces.
The ancilla states are orthogonal, i.e. $\langle{\mathbf i}|
{\mathbf j} \rangle =\delta_{ij}$. The vectors
$|a_i \rangle, |b_i \rangle$ denote the Schmidt bases of $|\psi \rangle$ 
and $u:= \floor\left( n/k\right)k$ where floor indicates the integer part \cite{order}.
 We denote the action of this map as ``{\em lifting up}''. In the following we describe the properties of this map
and its action on states and operators.

\vspace*{12pt}
\noindent {\bf Remark 2:}
For each pure state $|\psi\rangle = \sum_{i=1}^n \lambda_i |a_i b_i\rangle \in {\cal H}_A \otimes {\cal H}_B$ it holds that:
\begin{equation}
I_{k}(|\psi\rangle)= I_{k}(\sum_{i=1}^{k} \lambda_i |a_i b_i\rangle )  
+ I_{k}(\sum_{i=k+1}^{2k} \lambda_i |a_i b_i\rangle )  +... +
I_{k}(\sum_{i=u+1}^{n} \lambda_i |a_i b_i\rangle ).
\end{equation}
In particular, $I_{k}$ maps by definition pure states in ${\cal H}_A \otimes {\cal H}_B$
with Schmidt rank  $k$ or less into pure product states in $\calAB{k}$.\\
 
We now define for each operator acting on $\calab$
an operator acting on \linebreak[4] $\calAB{k}$.

\vspace*{12pt}
\noindent {\bf Definition 4:}
For a given  $S= \sum_{i,j,l,m} \sigma_{i,j,l,m} |i\rangle\langle l|_A \otimes
|j \rangle \langle m|_B$ acting on ${\cal H}_A \otimes {\cal H}_B$, and 
$k \in {\mathbb N}$ we define
an operator ${\mathbb S}_k$ on $\calAB{k}$ by:
\begin{equation}
{\mathbb S}_k:=\sum_{{\mathbf s,\mathbf t }=1}^k\sum_{i,j,l,m}
\sigma_{i,j,l,m} |i, \mathbf s \rangle \langle l, \mathbf t| \botimes
|j,\mathbf s \rangle \langle m, \mathbf t|.
\end{equation}
If 
 $\Tr(S)=1$, then
$\Tr({\mathbb S}_k)=k$.

\vspace*{12pt}
\noindent {\bf Remark  3:}
Reordering  the tensor-product structure of the enlarged Hilbert space 
as \linebreak[4]
\mbox{$\mathcal H_A \otimes \mathcal H_B \otimes \field^n \otimes \field^n$}
leads to the following expression for ${\mathbb S}_k$:
\begin{equation} 
\label{sdef}
{\mathbb S}_k = 
\sum_{{\mathbf s,\mathbf t }=1}^k S \otimes |\mathbf{ss} 
\rangle \langle \mathbf{tt}|.
\end{equation}
In the following the notation ${\mathbb S_k}$  (i.e. blackboard font)
will be used only for those operators acting on $\calAB{k}$, which can be
written like in eqn. (\ref{sdef}), with a corresponding operator $S \in
{\mathcal B}(\calab)$.

\vspace*{12pt}
\noindent{\bf Remark 4:}
For every pure state 
${\ket{\psi}} \in \calab$ with Schmidt rank  less or equal to  $k$, i.e.
$|\psi \rangle= \sum_{j=1}^{k} \lambda_i |a_i b_i \rangle$ with
$\lambda_i \geq 0$
and every operator $S$ it holds that:
\begin{eqnarray}
\langle \psi | S | \psi \rangle &=&
\sum_{s,t=1}^{k} \lambda_s^* \lambda_t \langle a_s b_s | S| a_t b_t
\rangle \nonumber \\
&=&
\sum_{i,j,l,m,s,t=1}^{k} \lambda_j^* \lambda_m \langle a_i | \langle b_j | 
S | a_l \rangle | b_m \rangle \langle \mathbf{ij}|\mathbf{ss} \rangle 
\langle \mathbf{tt} | \mathbf{lm} \rangle\nonumber \\
&=&
\langle I_{k}( \psi ) | {\mathbb S}_{k} | I_{k}(\psi ) \rangle.
\end{eqnarray}

\vspace*{12pt}
\noindent {\bf Remark 5:}
Notice that the same construction holds for a pure state 
of Schmidt rank larger than $k$, namely
$|\psi \rangle = \sum_{i=1}^{n} \lambda_i |a_i \rangle
|b_i \rangle=\sum_{j=1}^u |\psi_j \rangle$, 
with $u={\floor(n/k)}$ and  the abbreviation
$|\psi_j \rangle = \sum_{l=(j-1)k+1}^{\min(jk,n)} 
\lambda_l |a_lb_l \rangle$.  
By using Remark 2 one immediately obtains:
\begin{eqnarray}
\langle \psi | S | \psi \rangle 
&=& \sum_{i,j=1}^{u} \langle \psi_i | S | \psi_j \rangle
=\sum_{i,j=1}^{u} \langle I_{k}(\psi_i) | {\mathbb S}_{k} | I_{k}(\psi_j) 
\rangle \nonumber \\
&=& \langle I_{k}( \psi ) | {\mathbb S}_{k} | I_{k}(\psi ) \rangle.
\end{eqnarray}

\vspace*{12pt}
\noindent{\bf Remark 6:}
Given a mixed state $\rho$ and an arbitrary decomposition 
$\rho=\sum_{i=1}^{l} p_i |\phi_i \rangle \langle \phi_i|$, it follows that 
\begin{equation}
\Tr(S\rho) = \sum_{i=1}^{l} p_i \langle \phi_i|S|\phi_i\rangle \\
=\sum_{i=1}^{l} p_i \langle I_{k}(\phi_i) |{\mathbb S}_{k} | I_{k}(\phi_i) 
\rangle.
\end{equation}
\noindent Thus, by  defining a (non-normalized) mixed state 
$\Gamma_k:=\sum_i p_i
 |I_k(\phi_i) \rangle \langle I_k(\phi_i)| \in$\linebreak[4]${\cal B}(\calAB{k})$ one arrives at
\begin{equation}
\Tr(S\rho) = \Tr({\mathbb S}_k \Gamma_k).
\end{equation}

\subsection{
Mapping from the enlarged Hilbert space to the original
one}

In this part we now define  a map which transforms states in the
enlarged Hilbert space into states in the original one. 

\vspace*{12pt}
\noindent{\bf Remark 7:}
Every pure product state  $|A\rangle \botimes |B \rangle=|A,B\rangle \in \calAB{k}$ can be
expressed by using the Schmidt decomposition of each pure state $|A\rangle, |B\rangle$
in the bipartite splitting ${\mathcal H}_{A,B} \otimes \field^k$ as:

\begin{eqnarray}
|A \rangle = \sum_{i=1}^{k} \lambda_i |a_i \rangle \otimes |c_i \rangle
 &; &
|B \rangle = \sum_{j=1}^{k} \mu_j |b_j \rangle \otimes |d_j \rangle ,
\end{eqnarray}
where 
the Schmidt coefficients $ \lambda_i$ and $\mu_j$ are positive and
$\sum_i\lambda_i^2=1, \sum_j\mu_j^2=1$. 

\vspace*{12pt}
\noindent{\bf Definition 5:}
By $\tilde{J}_{k}: \calAB{k} \rightarrow \calab$ we denote the map that
transforms any pure product state 
$|A \rangle \botimes | B \rangle\in \calAB{k}$ into a pure 
state $\ket{\psi}\in {\cal H}_A \otimes {
\cal H}_B$ with Schmidt rank less or equal to $k$. This map is defined by:
\begin{eqnarray}
\label{tildj}
\tilde{J}_{k}: |A \rangle \otimes | B \rangle \mapsto 
| \tilde{J}_{k}(A \otimes B) \rangle &=&
\sum_{i=1}^k \langle \mathbf{ii}|
\left(\sum_{l=1}^{k} \lambda_l |a_l \rangle \otimes |c_l \rangle\right) \otimes
\left(\sum_{m=1}^{k} \mu_m |b_m \rangle \otimes |d_m \rangle\right) \nonumber\\
&=&\sum_{l,m=1}^{k} F_{lm} \lambda_l \mu_m |a_l b_m \rangle 
\end{eqnarray}
where $F_{lm}:=\sum_{i=1}^{k} \langle \mathbf{ii} | c_l d_m
\rangle$. 

Recalling that the Schmidt rank of a pure state is equal to the rank of
its reduced density matrices we find:  
\begin{eqnarray}
\Tr_B |\tilde {J}_{k}(A \otimes B) \rangle \langle\tilde{J}_{k}(A \otimes B) | 
&= &\sum_{s=1}^n \langle 
b_s | \left[
\sum_{i,j,l,m=1}^{k} F_{ij}^* F_{lm} \lambda_i^* \lambda_l \mu_j^*
\mu_m |a_l \rangle \langle a_i | \otimes |b_m \rangle \langle b_j | \right]
| b_s \rangle
\nonumber \\
&= &\sum_{i,l=1}^{k} \lambda_i^* \lambda_l \sum_{s=1}^{k} F_{is}^* F_{ls}
|\mu_s|^2  |a_l \rangle \langle a_i | .
\end{eqnarray}
The rank of this operator cannot exceed 
$k$, and therefore, the Schmidt rank of 
$|\tilde{J}_{k}(A \otimes B) \rangle$ does not exceed ${k}$. 

This map can be now straightforwardly extended to map 
entangled pure states of the enlarged space into the original one 
by using the Schmidt decompositions according to the split $\calAB{k}$ 
(Remark 7) 
and applying the map to each Schmidt term separately.

\vspace*{12pt}
\noindent {\bf Definition 6:}
By $J_{k}$ we define the extension of $\tilde{J}_{k}$ on all pure
entangled states $|\Psi \rangle$ in $\calAB{k}$:
\begin{equation}
J_k:|\Psi\rangle=\sum_{i=1}^{l} \lambda_i |A_i\rangle \botimes | B_i
\rangle 
\mapsto  
|J_{k}(\Psi)\rangle := \sum_{i=1}^l \lambda_i |\tilde{J}_{k} (A_i \otimes
 B_i ) \rangle.
\end{equation}
We call the action of the map $J_k$  ``{\em lifting down}''.

\vspace*{12pt}
{\bf Remark 8:}
Notice that for all $|\psi\rangle \in 
{\cal H}_A \otimes {\cal H}_B$ with Schmidt rank $k$ or less, it holds
that:
$\left| J_{k}\left( I_{k}(\psi)\right) \right\rangle=|\psi \rangle$. 
Thus, when  restricted to these states, the map
$J_{k}$ is the {\em inverse} map of $I_k$.\\

\subsection{Connection between the expectation values of operators }

After the definition of the two maps (the ``lifting up'' map $I_k$ and the
``lifting down'' map $J_k$) one observes that there is a close relation between 
the expectation value of an operator in the original space and the expectation value of
the corresponding operator in the enlarged space.

\vspace*{12pt}
{\bf Lemma 5:}
Given two arbitrary pure product states 
$|A_1 B_1 \rangle, |A_2 B_2\rangle\in$\linebreak[4]
$\calAB{k}$ the following equation:
\begin{equation}
\langle A_1 | \botimes \langle B_1| {\mathbb S}_{k} | A_2 \rangle \botimes 
|B_2 \rangle= \langle J_{k}(A_1 \botimes B_1) | S | 
J_{k} (A_2 \botimes B_2) \rangle
\end{equation}
holds.

\vspace*{12pt}
\noindent{\bf Proof:} 
We use Remark 3 to express ${\mathbb S}_{k}$ as a function of $S$,
 and express
each pure state   
$\ket {A_i}$
(respectively $\ket {B_i}$) in its 
Schmidt decomposition according to Remark 7. 
The expectation value of ${\mathbb S_k}$ is thus given 
 as:
\begin{eqnarray}
\langle A_1B_1 | {\mathbb S}_{k} |A_2B_2 \rangle 
&=& \left(\sum_{i,j=1}^{k} \lambda_i^* \mu_j^* 
\langle a_i b_j |S  (\sum_{l,m=1}^{k} \nu_l \xi_m |e_l f_m \rangle)\right) 
\sum_{s,t=1}^{k} \langle c_i d_j | \mathbf{ss} \rangle\langle \mathbf{tt}| g_l h_m \rangle 
\nonumber \\
&= &\left(\sum_{i,j=1}^{k} F_{ij}^* \lambda_i^* \mu_j^* \langle a_i b_j|\right)
 S \left(\sum_{l,m=1}^{k} G_{lm} \nu_l \xi_m |e_l f_m \rangle \right) 
 \nonumber \\ 
&= &\langle J_{k}(A_1 \botimes B_1)| S |(J_{k}(A_2 \botimes B_2) \rangle,
\end{eqnarray}
where 
$F_{ij}:= \sum_{s=1}^{k} \langle \mathbf{ss} | c_i d_j \rangle$ and
$G_{ij}:= \sum_{t=1}^{k} \langle \mathbf{tt} | g_i h_j \rangle$ $\square$\,. \\

\noindent So far, we have defined the action of the maps on pure states and operators and we have shown how
the maps permit to  ``jump"  from the original space to the enlarged one (and vice versa). We have  also shown 
the relation between the expectation value of an operator in the original space and the corresponding operator in the enlarged space. We proceed now to show  
that a Schmidt number witness ($k$-SW) acting in $\calab$ 
corresponds to an entanglement witness (2-SW)  acting on 
$\calAB{k}$. The main results of our paper are stated in the following
two theorems.

\vspace*{12pt}
\noindent {\bf Theorem 2:}\\
\noindent i)
Given two arbitrary operators $S,\rho \in {\cal B}(\calab)$ and 
an arbitrary decomposition \linebreak[4]
\mbox{$\rho=\sum_{i=1}^l p_i  
|\phi_i\rangle \langle \phi_i|$},
it holds that 
$\Tr(S \rho) = \Tr({\mathbb S}_k {\Gamma}_k)$, where 
${\Gamma}_k:= \sum_{i=1}^l p_i |I_k(\phi_i) \rangle \langle I_k(\phi_i)|$.\\
\noindent ii) 
Given two arbitrary operators ${\mathbb S}_k, \Theta \in {\cal
  B}(\calAB{k})$ and an arbitrary decomposition 
$\Theta=\sum_i p_i|\Phi_i \rangle
\langle \Phi|$, 
it holds that
$ \Tr( {\mathbb S}_k \Theta) = \Tr( S \theta)$, where 
$\theta:=\sum_i |J_k(\Phi_i)\rangle \langle J_k(\Phi_i)|$.

\vspace*{12pt}
\noindent{\bf Proof:} (i) See Remark (6). (ii) The proof is a concatenation of the previous remarks $\square$\,.

\vspace*{12pt}
\noindent {\bf Theorem 3:}\\
i) If $S$ is a $k$-SW acting on ${\cal H}_A \otimes {\cal H}_B$, then 
${\mathbb S}_{k-1}$ is a
$2$-SW acting on $\calAB{k-1}$.\\
ii) If ${\mathbb S}_{k-1}$ is a $2$-SW, 
then $S$ is an $n$-SW with  $k \leq n \leq 2k$. \\
iii) If ${\mathbb S'}_{k-1}= (1+\epsilon) {\mathbb S}_{k-1} - \epsilon
{\mathbb Z}_{k-1}$ is a  2-SW which is 2-finer than ${\mathbb S}_{k-1}$, 
then the corresponding $S'$ is $k$-finer than $S$.

\vspace*{12pt}
\noindent{\bf Proof:} (i) According to Remark 4 
and to the fact that $S$ is a $k$-SW one observes  
that ${\mathbb S}_{k-1}$ has a positive expectation value 
for all pure product-states in $\calAB{k-1}$. It remains to be
shown that there exists a state with Schmidt rank 2 in 
$\calAB{k-1}$
for which ${\mathbb S}_{k-1}$ has a negative expectation value. 
Since $S$ is a $k$-SW there exists a state
$|\psi \rangle$ with Schmidt rank $k$ that is detected by $S$. The 
Schmidt decomposition of such a state can be written as:
\begin{equation}
\ket{\psi}   = \sum_{i=1}^k \lambda_i |a_i \rangle |b_i \rangle.
\end{equation}
Notice then that $I_{k-1}$ will map this pure state of Schmidt rank $k$ 
into a pure state of Schmidt rank  $2$. 
According to Remarks 2 and 4
the expectation value of 
$\langle I_{k-1} (\psi) |{\mathbb S}_{k-1}|I_{k-1}(\psi) \rangle =
\langle \psi | S| \psi\rangle<0$. So a state with Schmidt rank $2$ 
in $\calAB{k-1}$
is
detected and ${\mathbb S}_{k-1}$ is a (non-normalized) 2-SW.

ii) Let ${\mathbb S}_{k-1}$ be a (non-normalized) 2-SW, then there exists a 
state 
$|\Psi \rangle :=  \kappa_1 |A_1 \rangle \otimes |B_1 \rangle + \kappa_2 | A_2 \rangle 
\otimes |B_2 \rangle$ with $\langle \Psi | {\mathbb S}_{k-1} 
|\Psi \rangle <0$, but since ${\mathbb S}_{k-1}$ is a 2-SW it is 
non-negative on all separable 
states in  $\calAB{k-1}$.

Since all states of Schmidt rank  less than $k$ are mapped by $I_{k-1}$ to product states,
one obtains that $S$ is non-negative on all states with 
Schmidt rank  less than $k$. Furthermore $S$ cannot be positive, 
since ${\mathbb S}_{k-1}$ is not positive. So it remains to 
be shown that there exists a state with Schmidt rank $n$ 
(with $k \leq n \leq 2k$) that is detected. Writing the pure states  
$|A_i\rangle$, $|B_i\rangle$ in their Schmidt decomposition as 
 in eqn. (17), the expectation value of  ${\mathbb S}_{k-1}$ is given
 as:

\begin{align*}
\langle \Psi | {\mathbb S}_{k-1}
| \Psi \rangle &=
|\kappa_1|^2 \langle A_1 B_1 |{\mathbb S}_{k-1}| A_1B_1\rangle + |\kappa_2|^2
\langle A_2 B_2 | {\mathbb S}_{k-1}| A_2B_2 \rangle \\
&+ \kappa_1^* \kappa_2 \langle A_1B_1| {\mathbb S}_{k-1}| A_2B_2 \rangle +
\kappa_2^* \kappa_1 \langle A_2 B_2 | {\mathbb S}_{k-1} | A_1 B_1 \rangle.
\end{align*}
By use of Lemma 5 one can relate each of  the above terms to 
a matrix element of $S$ and arrives at:
\begin{equation*}
\langle \Psi | {\mathbb S}_{k-1} | \Psi \rangle =
(\kappa_1^* \langle J_{k-1}(A_1 \otimes B_1 ) | + \kappa_2^*
\langle J_{k-1}(A_2 \otimes B_2)|) S (
(\kappa_1 | J_{k-1}(A_1 \otimes B_1 ) \rangle  + \kappa_2 | J_{k-1}(A_2 \otimes B_2)
\rangle).
\end{equation*}
Since $| J_{k-1}(A_1 \otimes B_1 ) \rangle$ and $| J_{k-1}(A_2 \otimes B_2)
\rangle$ both have  a Schmidt rank less than $k$, the Schmidt rank of their
sum cannot exceed $2 (k-1)$. So there exists also a minimal $n$ with
$k \leq n \le 2 (k-1)$ and a pure state in ${\cal H}_A \otimes
{\cal H}_B$ with Schmidt rank $n$, that is detected by $S$.\\
\noindent iii)
Let 
$
|\phi \rangle = \sum_{i=1}^k \lambda_i |a_i b_i \rangle =
|\psi \rangle + \lambda_k |a_k b_k \rangle
$
be an arbitrary 
pure state with Schmidt rank less or equal $k$ in $\calab$,
i.e. $\lambda_i \geq 0$.
Since ${\mathbb S'}_{k-1}$ is 2-finer than ${\mathbb S}_{k-1}$ the operator 
${\mathbb Z}_{k-1}$ has to be non-negative on all pure states  
$|\Psi \rangle = \mu_1 |A_1 B_1 \rangle + \mu_2 |A_2 B_2 \rangle$
with Schmidt rank two or less
in $\calAB{k-1}$, i.e. $\mu_{1,2} \geq 0$.
In particular this has to hold for 
$\mu_1| A_1 B_1 \rangle = I_{k-1} 
( |\psi \rangle )$ and $\mu_2|A_2 B_2 \rangle = 
I_k(\lambda_k |a_k \rangle \otimes |b_k \rangle)= 
|a_k \rangle \otimes |{\bf 1}\rangle \otimes  \lambda_k |b_k \rangle \otimes 
|{\bf 1}\rangle$. By calculating the expectation value of ${\mathbb Z}_{k-1}$ 
on $|\Psi \rangle$ and using Lemma 5 and Remark 8 one obtains
\begin{align*}
0 \leq 
\langle \Psi | {\mathbb Z}_{k-1} | \Psi \rangle &=
\langle I_k(\psi) | {\mathbb Z}_{k-1} | I_k(\psi) \rangle +
\langle I_k(\psi) | {\mathbb Z}_{k-1} | I_k(\lambda_k a_k \otimes b_k) \rangle
\\
&+
\langle I_k(\lambda_k a_k \otimes b_k) | {\mathbb Z}_{k-1} | I_k(\psi) \rangle 
+\langle I_k(\lambda_k a_k \otimes b_k) | {\mathbb Z}_{k-1} | 
I_k(\lambda_k a_k \otimes b_k) \rangle \\
&=\langle \phi | Z | \phi \rangle.
\end{align*}
Therefore, $ Z \in {\cal Q}_{(SN)_k}$, and due to Lemma 2 
$S'$ is $k$-finer than $S$ $\square$\,.

\vspace*{12pt}
\noindent{\bf Remark 9:}
An operator $S \in {\mathcal B}(\calab)$ is a $k$-SW iff 
the operator ${\mathbb S}_l$ is a (non-normalized) 
entanglement witness for all $l \leq k$, but no entanglement witness
for all $l >k$. 

\vspace*{12pt}
\noindent {\bf Lemma 6:} 
If there exists some $Z \in {\cal Q}_{(SN)_k}$ such that 
$Z P_{S^{k \times k}} =0$ and
\begin{eqnarray}
\label{lambda0}
\lambda_0 &= &\inf_{|A\rangle \in {\bf H}_A\otimes \field^{k-1}} \left[
(\langle A|{\mathbb Z}_{k-1}|A \rangle)^{-1/2} 
\langle A |{\mathbb S}_{k-1}|A\rangle (\langle A|{\mathbb Z}_{k-1}| A \rangle)^{-1/2} 
\right]_{\min} \\\nonumber
&=& \left(\sup_{|A \rangle \in {\bf H}_A \otimes \field^{k-1}} \left[ 
(\langle A|{\mathbb S}_{k-1}|A \rangle)^{-1/2}  \langle A|{\mathbb Z}_{k-1}| A\rangle
(\langle A|{\mathbb S}_{k-1}|A\rangle)^{-1/2}\right]_{\max}\right)^{-1} >0\ ,
\end{eqnarray}
where we denote by $[...]_{\min/\max}$ the minimal/maximal
 eigenvalue of an operator, then the operator
\begin{equation*}
S'(\lambda):= (S-\lambda Z)/(1-\lambda)
\end{equation*}
with $\lambda \geq 0$ is a $k$-SW if and only if $\lambda \leq \lambda_0$.

\vspace*{12pt}
\noindent{\bf Proof:}
Let us find out for which values of $\lambda >0$ the operator 
$S'(\lambda)$ is a $k$-SW and, therefore, 
${\mathbb S}'_{k-1}(\lambda )$ is an entanglement witness. 
To this aim we demand that 
 $\langle A | {\mathbb S}'_{k-1}(\lambda) | A \rangle
\geq 0$, i.e.
\begin{equation}
\label{bigsprime}
\langle A |{\mathbb S}_{k-1}|A \rangle 
- \lambda \langle A |
{\mathbb Z}_{k-1} | A \rangle \geq 0.
\end{equation}
On one hand, multiplying equation (\ref{bigsprime}) from the left and from the right 
by $(\langle A|{\mathbb Z}_{k-1}|A \rangle)^{-1/2}$ 
we obtain
$(\langle A|{\mathbb Z}_{k-1}| A \rangle)^{-1/2}\langle A |{\mathbb S}_{k-1}
| A \rangle (\langle A{\mathbb Z}_{k-1}| A \rangle )^{-1/2} \geq \lambda 
\eins$, which leads to $\lambda \leq \lambda_0$ given in 
the first part of the eqn. (\ref{lambda0}).
On the other hand, multiplying equation (\ref{bigsprime}) 
by $(\langle A|{\mathbb S}_{k-1}|A\rangle)^{-1/2}$ from 
the right and the
left side we obtain $(\langle A|{\mathbb S}_{k-1}| A\rangle)^{-1/2}
\langle A|{\mathbb Z}_{k-1}|A\rangle(\langle A|{\mathbb S}_{k-1}|A\rangle
)^{-1/2} \leq  \eins/\lambda $, which
 immediately leads to
$\lambda \leq \lambda_0$, given
in  the second equality of eqn. (\ref{lambda0}) $\square$\,.

\subsection{Example}
The aim of this subsection is to illustrate the previous method and 
results with
an explicit example. Consider the following one parameter family 
of operators
\begin{equation}
S(a):=\frac{1}{1-a}\left( \frac{1}{9} \eins -a |\psi \rangle \langle 
\psi|\right)
\end{equation}
acting on $C^3\otimes C^3$ where $|\psi\rangle= \frac{1}{\sqrt{3}} ( |00 \rangle + |11\rangle
+|22\rangle)$ and $a>0$.

Our goal is to determine for which parameters $a$ the witness
operator $S(a)$
is 
able to detect Schmidt number 3 only, i.e. for which it is
non-negative on states with Schmidt number 2. Note that the partial transpose
 of $S(a)$ provides a family of states that for some $a$ are 
$n$-copy non-distillable \cite{conjec}. 
In \cite{conjec}, where the possibility 
of the existence of non-distillable states with non-positive partial transpose
was discussed,
these states were investigated, and the result that we are
going to derive below, was obtained by using a direct method. 
The aim of the example presented 
here is thus to illustrate
how one can arrive at such result by transforming the problem to the 
task of checking if in some extended space a corresponding new operator 
is an  entanglement witness.

Notice that $S(a)$ is positive semidefinite and, therefore, not a witness if $a \leq \frac{1}{9}$. 
For an arbitrary $|e\rangle \in {\cal H}_A$ with $|e\rangle=\lambda_0
|0\rangle + \lambda_1 |1\rangle +\lambda_2 |2 \rangle$ the expectation value of $S(a)$ becomes:
\begin{align*}
S(a)_e :=
(1-a) \langle e | S(a)| e\rangle = &\frac{1}{9} \langle e| \eins_A |e
\rangle \eins_B - a \langle e | \psi \rangle \langle \psi |
e \rangle \\ 
 = &\frac{1}{9} \eins_B -\frac{a}{3}
  ( |\lambda_0|^2 |0 \rangle \langle 0 |
+ \lambda_0^* \lambda_1 |0 \rangle \langle 1| 
+ \lambda_0^* \lambda_2 |0 \rangle \langle 2| 
+ \lambda_1^* \lambda_0 |1 \rangle \langle 0| \\ 
&+ |\lambda_1|^2 |1\rangle \langle 1 | 
+ \lambda_1^* \lambda_2 |1 \rangle \langle 2| 
+ \lambda_2^* \lambda_0 |2 \rangle \langle 0| 
+ \lambda_2^* \lambda_1 |2 \rangle \langle 1| 
+ |\lambda_2 |^2 |2 \rangle\langle 2| ) \\
= &\left(\begin{array}{ccc}
\frac{1}{9}-\frac{a}{3}|\lambda_0|^2 & -\frac{a}{3}\lambda_0
\lambda_1^* & -\frac{a}{3}\lambda_0 \lambda_2^* \\ 
-\frac{a}{3} \lambda_1 \lambda_0^* & \frac{1}{9}-\frac{a}{3}
|\lambda_1|^2 & -\frac{a}{3} \lambda_1 \lambda_2^* \\ 
-\frac{a}{3} \lambda_2 \lambda_0^* &-\frac{a}{3} \lambda_2 \lambda_1^*
& \frac{1}{9}-\frac{a}{3}|\lambda_2 |^2  
\end{array}
\right).
\end{align*}
The operator
$S(a)_e$ has the eigenvalues $\{ \frac{1}{9}, \frac{1}{9}, \frac{1}{9}- 
\frac{a}{3} \}$. Therefore, by definition $S(a)_e$ 
is an entanglement witness  for $\frac{1}{9} < a < \frac{1}{3}$. 
Does exist, however, a region of the parameter space  for which 
$S(a)$ is a 3-SW, i.e. it detects a state with Schmidt rank 3 but does
{\em  not}
detect any state with Schmidt rank 2?
Clearly for all $\frac{1}{9} < a\leq
\frac{1}{3}$ where $S(a)$ is no $3$-SW it is  a $2$-SW.

According to Theorem 3 an operator $S(a)$ on a $3 \times
3$-dimensional Hilbert-space is a 3-SW iff 
${\mathbb S}_2(a):=S(a) \otimes 
(|\mathbf{00}\rangle +|\mathbf{11} \rangle ) (\langle \mathbf{00}| + \langle 
\mathbf{11}|)$ is
a (non-normalized) 2-SW. This new operator fulfils 
that for all
pure states $|E \rangle \in {\mathcal H}_A \otimes {\mathcal C}^2$ 
the operator $\langle E | {\mathbb S}_2(a)|E\rangle \geq 0$. Furthermore 
since any pure state $|E\rangle$ can be decomposed in an arbitrary basis 
$\{|1\rangle,|2\rangle,|3\rangle\}$ of ${\mathcal H}_A$ as
$|E\rangle= \mu_0|0, \mathbf0 \rangle + \mu_1 |0, \mathbf 1 \rangle +\mu_2
|1, \mathbf 0 \rangle + \mu_3 | 1, \mathbf 1 \rangle + \mu_4 | 2, \mathbf 0 
\rangle + \mu_5 |2, \mathbf 1\rangle$, the above operator can be expressed as:
\begin{align*}
\hspace{-1.5cm}
\langle E | {\mathbb S}_2(a)|E\rangle = \frac{1}{1-a} 
\left(
\begin{array}{cccccc}
A -\frac{1}{3}a |\mu_0|^2 & 
B- \frac{1}{3}a \mu_0^*\mu_1&
-\frac{1}{3} \mu_0^* \mu_2 &
-\frac{1}{3} \mu_0^* \mu_3 &
-\frac{1}{3} \mu_0^* \mu_4 & 
-\frac{1}{3} \mu_0^* \mu_5  \\
C- \frac{1}{3}a \mu_1^*\mu_0 &
D-\frac{1}{3}a |\mu_1|^2 &
-\frac{1}{3} \mu_1^* \mu_2 &
-\frac{1}{3} \mu_1^* \mu_3 &
-\frac{1}{3} \mu_1^* \mu_4 &
-\frac{1}{3} \mu_1^* \mu_5 \\
-\frac{1}{3} \mu_2^* \mu_0 &
-\frac{1}{3} \mu_2^* \mu_1 &
A -\frac{1}{3}a |\mu_2|^2 &
B- \frac{1}{3}a \mu_2^*\mu_3 &
-\frac{1}{3} \mu_2^* \mu_4 &
-\frac{1}{3} \mu_2^* \mu_5 \\
-\frac{1}{3} \mu_3^* \mu_0 &
-\frac{1}{3} \mu_3^* \mu_1 &
C- \frac{1}{3}a \mu_3^*\mu_2 &
D-\frac{1}{3}a |\mu_3|^2 &
-\frac{1}{3} \mu_3^* \mu_4 &
-\frac{1}{3} \mu_3^* \mu_5 \\
-\frac{1}{3} \mu_4^* \mu_0 &
-\frac{1}{3} \mu_4^* \mu_1 & 
-\frac{1}{3} \mu_4^* \mu_2 &
-\frac{1}{3} \mu_4^* \mu_3 &
A-\frac{1}{3}a |\mu_4|^2 &
B- \frac{1}{3}a \mu_4^*\mu_5 \\ 
-\frac{1}{3} \mu_5^* \mu_0 &
-\frac{1}{3} \mu_5^* \mu_1 &
-\frac{1}{3} \mu_5^* \mu_2 &
-\frac{1}{3} \mu_5^* \mu_3 &
C- \frac{1}{3}a \mu_5^*\mu_4 &
D-\frac{1}{3}a |\mu_5|^2 
\end{array}
\right)
\end{align*}
with $A=(\frac{1}{9}(|\mu_0|^2+|\mu_2|^2+|\mu_4|^2)$, 
$B=(\frac{1}{9}\mu_0^* \mu_1 + \mu_2^* \mu_3 +\mu_4^*\mu_5)$,
$C=(\frac{1}{9}\mu_1^* \mu_0 + \mu_3^* \mu_2 +\mu_5^*\mu_2)$
and 
$D=\frac{1}{9} (|\mu_1|^2+|\mu_3|^2+ü |\mu_5|^2)$.
It is tedious but straightforward to check that this operator is positive definite for $\frac{1}{6} \geq a$. 
Thus, we obtain that
\begin{equation*}
S(a) \mbox{ is } \left\{ 
\begin{array}{lcc}
\mbox{ positive (no witness)} & & \frac{1}{9} \geq a \geq 0 \\
\mbox{ a 3-SW } & \mbox{for} &\frac{1}{6} \geq a > \frac{1}{9} \\
\mbox{ a 2-SW } & &\frac{1}{3} \geq a > \frac{1}{6} 
\end{array}
\right.  . 
\end{equation*}

\vspace*{1pt}\textlineskip    
\section{Conclusions}
\label{Sect4}
In this article we have first reviewed  some properties of general
witness operators, as well as their optimisation. 
We have then focussed  on Schmidt number witnesses for bipartite systems, i.e. 
those witness operators which are capable to detect the minimal number
of entangled degrees of freedom between both parties (their
Schmidt number). 
We have shown that it is possible to relate any
 witness operator for Schmidt number $k$ 
to a  witness of Schmidt number 2  
in an enlarged Hilbert space in such a way that
the original subset of states with Schmidt number $(k-1)$ corresponds to
the subset of separable states in the enlarged space. The fact that
one can establish this
correspondence  
between a $k$-Schmidt witness  in 
the original Hilbert space 
and a  
Schmidt witness of  number 2 (i.e. a general entanglement witness)
in an enlarged Hilbert space
substantially simplifies the construction and optimization of the desired
$k$-Schmidt  witness. The reason for this is the fact that 
it is, in general, a much easier task  to check whether an operator 
is positive semidefinite on pure product states, rather than 
to check positivity on pure states of a
given Schmidt rank larger than one.  
Nevertheless a word of caution is needed when using this method for
optimization purposes only, as 
the concept of ``being finer'' is not generally preserved under the
lifting map. Therefore it is not always possible to optimize
a $k$-Schmidt witness by optimizing the corresponding 2-Schmidt witness
in the enlarged space.

\nonumsection{Acknowledgements}
\noindent
We acknowledge support from the ``Deutsche Forschungsgemeinschaft'' (DFG) via the
Schwerpunkt 1078 ``Quanten-Informationsverarbeitung'', the SFB 407 
``Quantenlimitierte Messprozesse mit Atomen, Molek\"ulen und Photonen'',
the
European Graduate College 665 ``Interference and Quantum Applications'' and 
the EU Programme QUPRODIS.

\nonumsection{References}


\begin{thebibliography}{99}
\bibitem{measurezero}
If $\cala$ is of measure 0 in $\calb$ and is convex there exists
a lower-dimensional subspace $\calb'$ of $\calb$ s.t. $\cala \subset
\calb'$ but not of measure 0. To check whether an element does not
belong to a subspace is fairly easy (by application of the 
projection onto that subspace). 
So without loss of generality $\cala$ is not of measure 0 in
$\calb$.
\bibitem{necess}
When the set 
$\cala$ is closed there exists, according to the Hahn-Banach theorem, 
for any $\rho \not \in \cala$
a linear functional $f$ on the
space of all bounded operators on ${\mathcal B}({\mathcal H})$, 
with the property that $f(\rho)<0$, but $f(\sigma) \geq 0$ for all $\sigma
\in \cala\mbox{}$. Furthermore, every linear
functional corresponds to a hermitian operator $W(f)$ defined through
$f(\rho)=\Tr( W(f)\rho)$.  Since $\eins \in \cala$ we know that $\Tr
(W(f)) >0$, therefore the operator $W(f)/(\Tr (W(f)))$ is a general
$({\cala,\calb})$-witness that detects $\rho$.
\bibitem{order}
The Schmidt decomposition is unique only up to a permutation of the terms. 
Thus, strictly speaking,  
the map $I_k$ is not defined unambiguously, 
due to different possible orderings
of the Schmidt terms. However, the  arguments
in the remainder
 of the paper are independent of this permutation dependence, and 
hold for any order of the positive Schmidt coefficients. 
Therefore, in the following we 
use for every  pure state $\ket{\psi}$ an arbitrary but fixed
order $|\psi\rangle = \sum_{i=1}^n \lambda_i |a_i \rangle \otimes 
|b_i \rangle$, where for all $i$ larger than the Schmidt rank of $\ket{\psi}$ 
we have $\lambda_i=0$.
\end{thebibliography}
\end{document}